\documentstyle[12pt,psfig]{article}
\title{An Age Difference of 2 Gyr between a Metal-Rich and a Metal-Poor Globular Cluster}

\author{ B. M. S.~Hansen$^1$, J. S. Kalirai$^{2,3}$, J. Anderson$^2$, A. Dotter$^4$, H. B. Richer$^5$,\\
 R. M. Rich$^1$, M. M. Shara$^6$, G. G. Fahlman$^7$, J. R. Hurley$^8$, I. R. King$^9$, \\
 D. Reitzel$^1$ \&  P. B. Stetson$^7$ }

\begin{document}

\maketitle

\begin{enumerate}
\item Division of Astronomy, University of California Los Angeles, Los Angeles, CA, 90095
\item Space Telescope Science Institute, 3700 San Martin Drive, Baltimore, MD 21218, USA
\item Center for Astrophysical Sciences, Johns Hopkins University, Baltimore, MD 21218
\item Research School of Astronomy and Astrophysics, Australian National University, Weston Creek, ACT 2611, Australia
\item Department of Physics \& Astronomy, University of British Columbia, Vancouver, BC, Canada
\item Department of Astrophysics, American Museum of Natural History, Central Park West and 79th Street, New York, NY 10024, USA
\item NRC, Herzberg, Victoria, BC, V9E 2E7, Canada
\item Center for Astrophysics \& Supercomputing, Swinburne University of Technology, Hawthorn, VIC 3122, Australia
\item Department of Astronomy, Box 351580, University of Washington, Seattle, WA 98195, USA

\end{enumerate}

\begin{abstract}
Globular clusters trace the formation history of the spheroidal components of both our Galaxy and others\cite{BS}, 
which represent the bulk of star formation over the history of the universe\cite{FHP}. They also exhibit a range of metallicities, 
with metal-poor clusters dominating the stellar halo of the Galaxy, and higher metallicity clusters found within the inner Galaxy, 
 associated with the stellar bulge, or the thick disk\cite{Zinn,Minn}. Age differences between these clusters can indicate the 
sequence in which the components of the Galaxy formed, and in particular which clusters were formed outside the Galaxy and later 
 swallowed along with their original host galaxies, and which were formed in situ. 
 Here we present an age determination of the metal-rich globular cluster 47~Tucanae by fitting the properties of the cluster white 
dwarf population, which implies an absolute age of $9.9 \pm 0.7$~Gyr at 95\% confidence. This is $\sim 2.0$~Gyr younger than inferred 
for the metal-poor cluster NGC~6397 from the same models, and provides quantitative evidence that metal-rich clusters like 47~Tucanae formed 
later than the metal-poor halo clusters like NGC~6397. 
\end{abstract}

The ages of globular clusters are most easily determined by fitting to the properties of main sequence and horizontal-branch stars\cite{SW98,Ros99,DeAng05,MF09,Dott11}, but these methods lead
to uncertainties caused by covariance between the measured age and the spectroscopic metallicity required to compare observed colours
and magnitudes to theoretical models. The metals in the atmospheres of white dwarfs sediment out because of the strong stellar gravity, resulting
in an age determination more robust with respect to metallicity correlations.
 In this case, the age is based on the monotonic cooling of white dwarfs,
such that the luminosity function peak moves to fainter magnitudes for older populations. The
consistency of the two approaches has been demonstrated with the determination of 
absolute ages for the globular clusters Messier~4 and NGC~6397\cite{Ha02,Ha04,Ha07,Bed09}. Our
attempt to extend the application of this method to the metal-rich cluster 47~Tucanae is 
driven by the fact that the 
covariance between metallicity and age in the main sequence fitting method gets stronger with
increasing metallicity, indicating the need for an independent determination.

We observed a field at roughly the half-mass radius of the cluster with the Hubble Space Telescope (HST).
 The primary science observations were performed with the Advanced Camera for Surveys (ACS), with
observations split between the wide F606W (total integration of 163.7 ks) and F814W (total integration of 172.8 ks) bandpasses.
The field was also chosen to overlap with prior HST observations, so that the cluster motion
results in positional shifts of cluster members
with respect to background stars, and so cluster and non-cluster members can be separated.
Details of the photometry, completeness and measurement of the proper motions is described in the Supplementary Information.

The white dwarf population is isolated by both colour and proper motion, as shown in Figure~1.
 The distribution of white dwarfs with luminosity (the luminosity function) can be compared (Figure~2)
 with the white dwarf luminosity function of another, metal-poor, globular
cluster with deep HST observations, NGC~6397\cite{Ha07}.
The abrupt truncation of these luminosity functions is a manifestation
of the finite age of a population, as white dwarfs cool monotonically, and this cutoff moves to fainter
magnitudes as a population ages. In Figure~2 we have adjusted the magnitudes of the NGC~6397  population
to correct for the different cluster distances, and so placed it on the same magnitude scale as the 47~Tuc
population. The 47~Tuc luminosity function drops to 50\% of the peak value at $\sim 0.4$~magnitudes brighter than that of
NGC~6397, a direct demonstration that this cluster is younger. A comparison to a third cluster, M4\cite{Bed09},
shows a similar age difference.

In order to determine a quantitative measure of the age difference, we construct Monte Carlo models for the cooling 
of the cluster white dwarf population, including a model for photometric shifts, scatter and incompleteness, based
on artificial star tests. We fit these models, including the uncertainties in distance and extinction, 
 to both the observed 
luminosity function and the two-dimensional distribution of the white dwarf population in colour
and magnitude (the Hess diagram). If we use the same models to fit the 47~Tuc data as we did for
NGC~6397\cite{Ha07}, we obtain an age of $9.7 \pm 0.4$~Gyr (at 95\% confidence) for 47~Tuc, as compared to the
corresponding value of $11.7 \pm 0.3$~Gyr for NGC~6397. Therefore, the relative age difference is $2.0 \pm 0.5$~Gyr.

The cluster 47~Tuc has a metallicity\cite{KMcW,McWB,Car09} of $\left[ \rm Fe/H \right]=-0.75$, compared to $\left[ \rm Fe/H \right]=-1.8$ for NGC~6397.
Although white dwarf atmospheres are not sensitive to overall metallicity, the cooling may be affected by changes in the nuclear
burning history due to different metallicities.
In order to quantify any difference introduced
by variations in progenitor behaviour, we have calculated new evolutionary models using progenitors of appropriate
metallicity for 47~Tuc. Using the MESA code\cite{Pax}, we have calculated the evolution of solar-mass stars
with $\left[ Fe/H \right] = -0.75$ and $\left[ \alpha / Fe \right]=+0.2$, as well as
with a range of Helium enrichment Y seen in 47~Tuc\cite{Milone}. We have also verified that these produce similar
results to the models used to determine the age of the cluster from the main sequence\cite{Dotter}, in order to facilitate comparison with those earlier analyses.
 We find that the final age determination
is not sensitive to either progenitor metallicity, Helium fraction or whether Schwarzschild or Ledoux criterion for convection is assumed when calculating
the mixing of the progenitor core. Neither is the result sensitive to the use of different atmospheric models for correcting
effective temperatures to observed colours. Details of this analysis are discussed in the Supplementary Information.
Marginalising over all these models results in an age of $9.9 \pm 0.7$~Gyr for 47~Tuc. Figure~3 shows the comparison between
the best fit model and the data, binned according to the grid in Figure~1.

Estimates based on fitting main sequence models to the main sequence turnoff report ages
between 10 and 13 Gyr, with the range due to a variety of factors
including differences in adopted models physics (such as nuclear reaction rates
and gravitational settling) as well as different bandpasses employed in the
observational data\cite{Grat03,Van2000,SW02,Sal07}.
To date, the most precise age obtained from the main sequence of 47 Tuc comes
from the detached, eclipsing binary star V69\cite{V69}.
V69 comprises two stars that lie just above and below the main sequence turnoff
meaning that the pair provides a particularly valuable age constraint.
The resulting age is somewhat sensitive to the assumed metallicity of the
stars. The original age estimate for V69\cite{V69} was
 $11.25 \pm 0.21 \pm 0.85$ Gyr, denoting random and
systematic errors, respectively. This result is based on the assumption
of [Fe/H] = -0.7 and [$\alpha$/Fe]=+0.4.
 Adopting a chemical composition in accordance with
more recent studies\cite{KMcW,McWB,Car09} and employing a more robust estimate of the systematic error
implies the age derived from the binary system is $10.39 \pm 0.54$ Gyr, here
citing only the systematic error. This estimate is
 fully consistent with the WD result given above,
suggesting that consistency between the two age measurement techniques
is possible, with accurate determination of extrinsic properties such
as distance, reddening and metallicity.

The bimodality in the metallicity distribution of globular clusters appears
to be ubiquitous among large galaxies\cite{BS}, and is held to reflect 
the sequence in which galaxies at high redshifts assemble their stellar
mass. It is still disputed whether the two classes represent two distinct
epochs of stellar assembly within a single dark matter halo, or whether some
clusters are assembled early in smaller haloes and then later accreted\cite{MG,SGF,Ton}.
Within this context, our result supports the idea that there is a measureable
age difference between metal-poor and metal-rich globular clusters in the Milky Way.
Figure~4 shows the relationship between
white dwarf-based age and metallicity for several Galactic populations. Including
the uncertainties due to stellar models, 47 Tuc has an absolute age $9.9\pm 0.7$~Gyr,
which places its formation at a redshift $\sim 2$, and consistent with the model
that the metal-rich globular clusters are formed in the massive, star-forming events
which are responsible for the peak in the global star formation rate at this redshift,
whether driven by mergers or in situ formation\cite{MG,SGF,Ton}.




{\bf Supplementary Information} is available in the online version of the paper.

Support for the program GO-11677 was provided by NASA 
through a grant from the Space Telescope Science Institute, 
which is operated by the Association of Universities for Re- 
search in Astronomy, Inc., under NASA contract NAS 5-26555.
This work is supported in part by the Natural Science and Engineering
Research Council of Canada.

 {\bf Correspondence:}  Correspondence and requests for materials
should be addressed to B.Hansen~(email: hansen@astro.ucla.edu).


\begin{figure}
\psfig{file=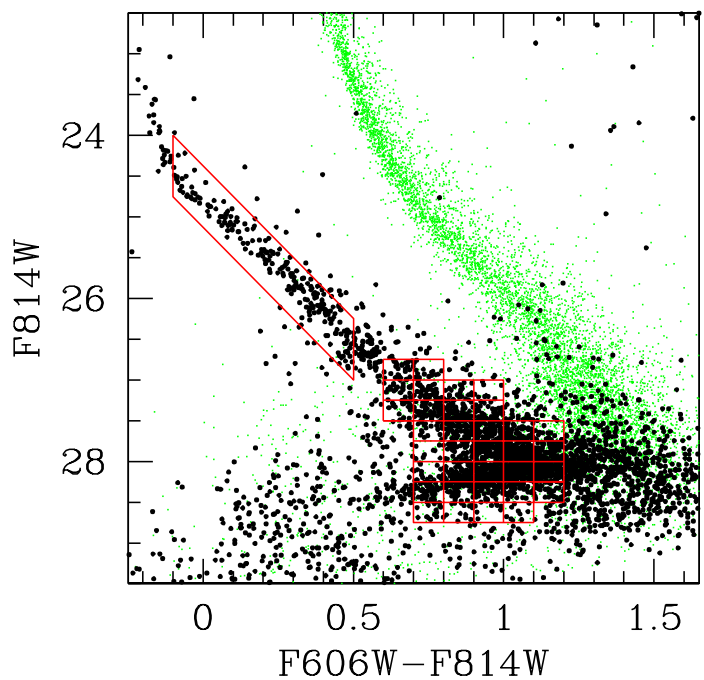,height=7in}
\caption{$\left| \right.$ {\bf The 47 Tucanae White Dwarf Cooling Sequence.} The filled black points
are the proper motion selected white dwarfs of 47~Tucanae, while the
small green points are those objects that fall outside the proper motion
cut. At brighter magnitudes (redder colours) these are members of the main
sequence population of the Small Magellanic Cloud, while they are mostly
background galaxies at fainter magnitudes (bluer colours).
 The grid outlined in red shows the binning used to compare the observations
to the Monte Carlo populations.}
\end{figure}

\begin{figure}
\psfig{file=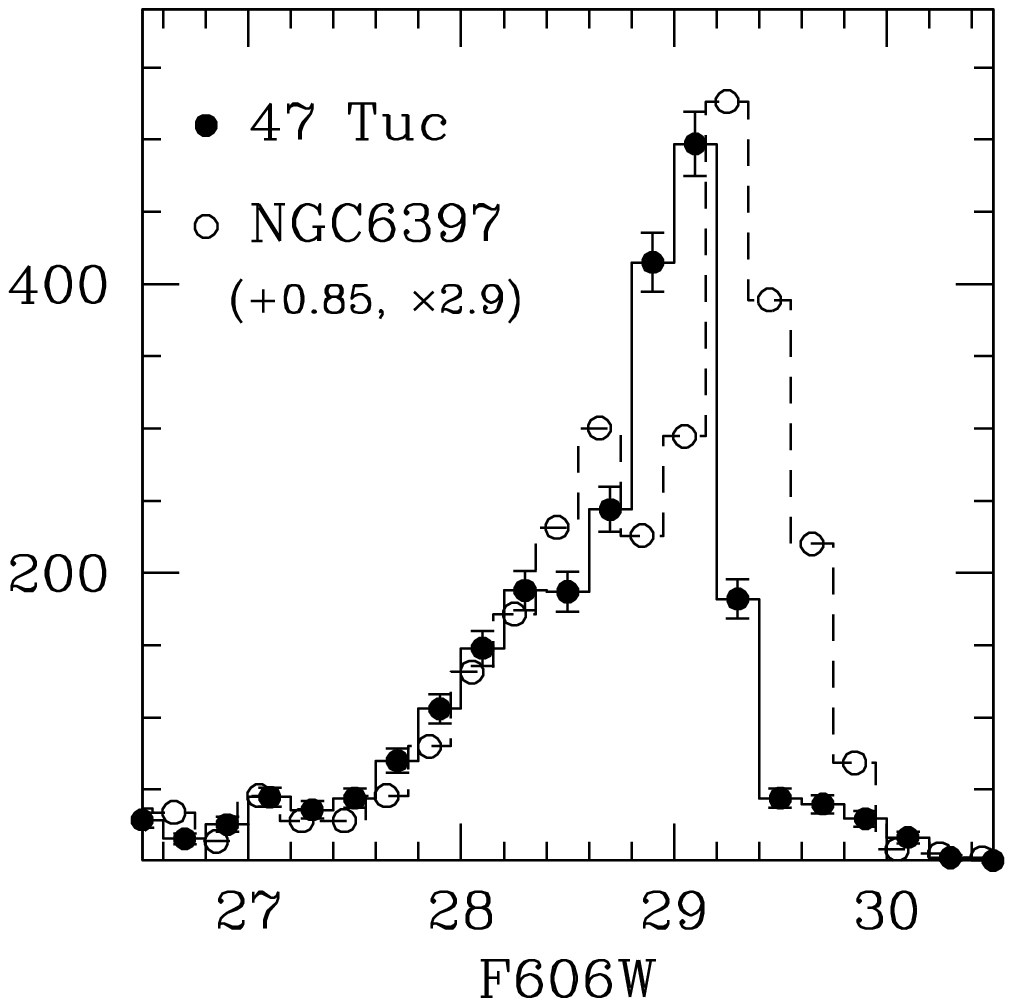,height=7in}
\caption{{\bf $\left| \right.$ Luminosity Function Comparison between 47~Tuc and NGC~6397.} The solid points and histogram show the white dwarf luminosity function for 47~Tuc, while
the
 open circles and dashed histogram is the corresponding quantity for NGC~6397\cite{Ha07}, with error bars based on Poisson statistics.
 Each shows
 the characteristic behaviour of populations that result from a single burst of star formation, namely
a sharp rise followed by a relatively abrupt truncation.
 In the case of NGC~6397 data, we have applied
a horizontal shift of 0.85 magnitudes to correct for the different cluster distances and put them on the same magnitude scale, and the
vertical scaling has been adjusted by a global multiplicative factor of 2.9 to correct for differences
in the absolute numbers of the different populations. We see that the 47~Tuc luminosity function clearly
truncates at a brighter luminosity than that of NGC~6397, indicating that
this cluster is manifestly younger.}
\end{figure}

\begin{figure}
\psfig{file=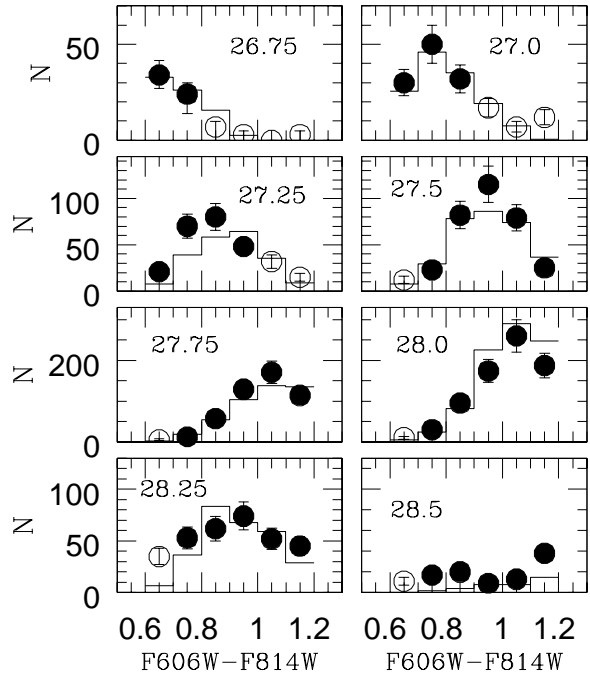,height=7in}
\caption{$\left| \right.$ {\bf Model Comparison to White Dwarf Colour-Magnitude Distribution.} In each panel, the points show the binning of the white dwarf population as a function of colour at fixed magnitude. Each panel is labelled with the corresponding central F814W magnitude.
 Filled circles are points that are included in the grid in Figure~1, while
open circles are not as they are expected to be dominated by residual contamination from background populations.
Error bars are Poisson errors.
The histogram is the model distribution binned in the same way, for an age of 9.7~Gyr and distance $\mu_{606W}=13.38$.}
\end{figure}

\begin{figure}
\psfig{file=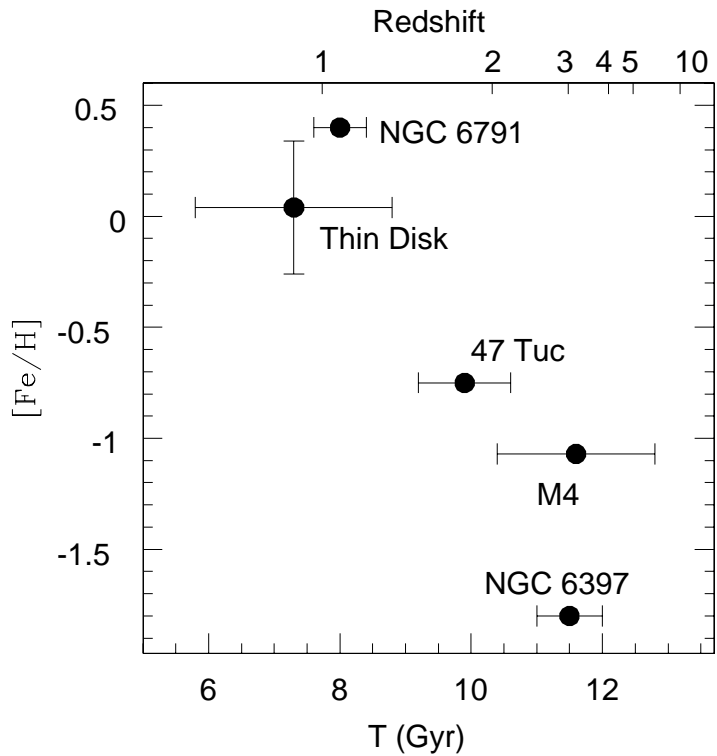,height=7in}
\caption{$\left| \right.$ {\bf Age-Metallicity Relation based on White Dwarfs.}
Each point shows the age of a Galactic population as determined using the white dwarf luminosity function,
and the corresponding metallicity determined from main sequence stars. The age of 47~Tuc is determined here, while those
 of NGC~6397\cite{Ha07},
 M4\cite{Bed09} and
the the open cluster NGC~6791\cite{GB10} are taken from the literature. The age of the
thin disk is taken from an analysis of the white dwarf population in the solar neighbourhood\cite{Ha02} and the metallicity error bar is determined from the
range of metallicities measured for nearby stars\cite{VF05}. The age uncertainties for all populations are quoted from
the given references.
 The top axis is labelled in redshift, using the age-redshift
relation appropriate to the WMAP cosmology\cite{W06}. This demonstrates that the age difference derived
here corresponds to the difference between $z\sim 3$ and $z\sim 2$.}
\end{figure}

\end{document}


\centerline{\bf Supplementary Information}

\bigskip

{\bf Deep Photometry in 47~Tuc }
\bigskip

Our data was taken with the ACS/WFC instrument on the Hubble Space Telescope.
This choice is dictated by the larger field of view and for ease of comparison
with prior epochs of observation in the same field, to facilitate proper motion
separation of cluster members. The wide F606W filter is the primary bandpass, given
its excellent throughput and a central wavelength well matched to the spectral
energy distribution of cluster white dwarfs. We also observed in a second, redder
filter (F814W), which allows us to probe the diminution of flux at longer wavelengths
in cool white dwarfs. The exposure time was split between 117 exposures in 
F606W and 125 deep exposures in F814W . The exposure
 times of the F606W frames ranged from 1113--1498 seconds
 for a total integration of 163.7 ks and 
those in F814W ranged from 1031--1484 seconds for an 
integration of 172.8 ks. The field was observed at multiple dither positions
and roll angles in order to correct for cosmic rays, diffraction spikes
and a variety of instrumental artifacts.

Given the complex design of the program, with multiple roll-angle restrictions, the
observations were collected over multiple epochs extending from January to 
October 2010. The calibrated files were produced with the {\it calacs} pipeline,
with additional processing to address the known CTE degradation in the ACS chip.

The primary goal of this study is to find and measure the faintest possible stars in the field. These often represent 
0.2$\sigma$ events relative to the background in individual images, and can only be seen by carefully analysing the full data set. Thus, 
the multiple images in a given filter were combined into a single coadded frame using the Multidrizzle algorithm$^{31}$, at a final supersampled grid with scale 0.03 arcsec/pixel.  This image (in each filter), which spans ~10,500 pixels in both dimensions, is well sampled and has a PSF with a FWHM$\sim$ 2.7 pixels.  

To measure the photometry, astrometry, and morphology of all sources on this image, the stand-alone versions of the DAOPHOT~II and ALLSTAR photometry programs were used$^{32,33}$.   An initial pass of 
DAOPHOT~II was performed to yield positions and aperture photometry of all possible detections that were at least 2.5$\sigma$ above the local sky in each of the two images for F606W and F814W.   Next, 1000 PSF candidate stars were selected based on brightness and isolation, and a PSF was calculated using an iterative method$^{34}$.  The final PSF allows for cubic variations with position in the frame.  The final step of the processing involves applying the PSF to the catalog of all sources in each image.  For this, ALLSTAR was used to perform both PSF-fitted 
astrometry and photometry, and also to retain morphological information of sources.

The photometric zero points are determined by measuring the brightnesses of isolated stars in the field on a single (distortion-corrected) image, and cross correlating these sources to our drizzled stack.  The final zero points, as measured from a few thousand stars, are 34.024 ($\sigma$ = 0.013) in F606W, and 33.222 ($\sigma$ = 0.014) in F814W$^{34}$.

\medskip

{\bf Artificial Star Tests}

\bigskip

In order to verify that the truncation of the white dwarf sequence is the result of finite population age and not systematic measurement effects, we need to convincingly demonstrate the completeness of our data. To that end we have performed a detailed set of artificial star tests, inserting objects of known brightness into our data and analysing them in the same manner as the real objects. This yields not only a recovery fraction, but also a scattering matrix that relates input and recovered magnitudes, and therefore measures the photometric shifts and scatter 
appropriate to our observations.

The artificial stars are modeled from the PSF of the drizzled stack, and scaled to reproduce the complete luminosity range of real stars in the data. The fraction of stars injected into each image was set to 5\% of the total number of stars in the image, so as to not induce incompleteness due to crowding in the tests themselves. One thousand trials were generated to form the input grid of artificial starlists and resulting images, with specific placement of the stars along the white dwarf cooling sequence. These images were then fed through the photometric routines that were applied to the actual drizzled images, using identical criteria. The stars were recovered blindly and automatically cross-matched to the input starlists containing actual positions and fluxes.

The magnitudes at which stars were recovered were recorded, so that we have not only a completeness fraction, but also a scattering matrix that related input magnitude to recovered magnitude. This is used in the model Monte Carlo simulations to simulate the photometric scatter. Artificial stars were inserted and recovered on the same grids in both F606W and F814W images, in order to account for potential correlations in the photometric scatter in the two bandpasses (because much of our background is due to stray light from bright stars in the field).

Figure~S1 shows the resulting recovery fraction in F606W for artificial stars inserted along the white dwarf cooling sequence, aligned with the observed white dwarf
luminosity function. We see that the observed peak of the white dwarf luminosity function occurs where completeness is still $\sim 70$\%. Also shown, as horizontal error bars, is the 1$\sigma$ dispersion in recovered magnitudes at each input magnitude, which is small enough to justify our choice of magnitude bin. Furthermore, the shift in the median recovered magnitude,
relative to input values, is $<0.04$ magnitudes down to F606W=29.4. This is well below the observed truncation and much smaller than the offset relative to NGC~6397.

\psfig{file=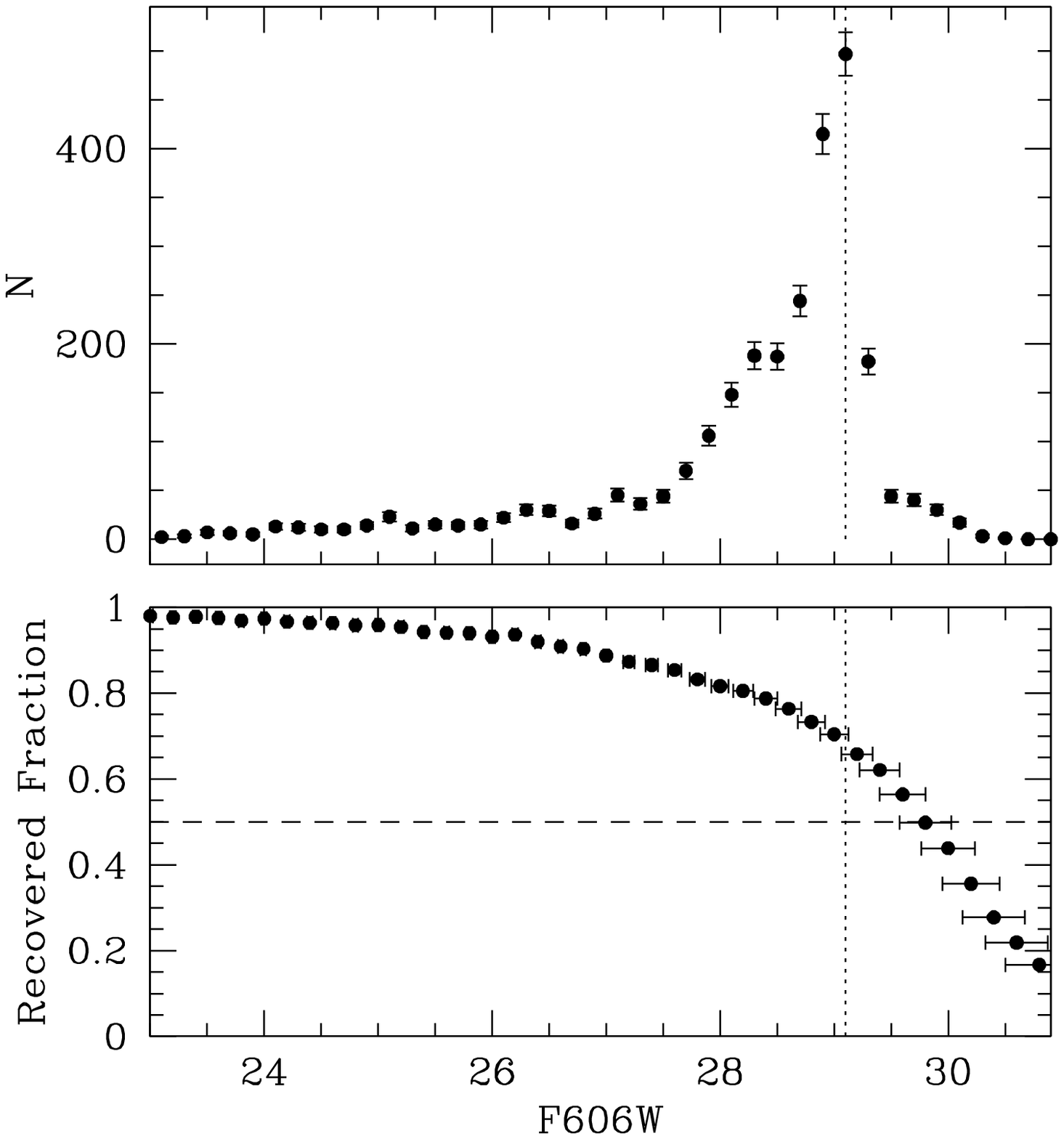,height=5in}

{\bf Figure~S1$\left| \right.$  Recovery Fraction for White Dwarf Stars}.
 The filled circles in the lower panel show the
fraction of successfully recovered artificial white dwarfs, as a function
of input F606W magnitude. The horizontal error bars indicate the dispersion
in the recovered magnitudes at each input value. In the upper panel we show
the luminosity function of 47~Tuc white dwarfs, on the same magnitude scale.
 The vertical dotted line in both panels
shows the magnitude at which the white dwarf luminosity function peaks, and
the horizontal dashed line indicates 50\% recovery fraction. Together these
indicate that the sharp drop in white dwarf counts is real and not a result
of incompleteness.

\medskip

{\bf Proper Motions}
\bigskip

Proper motions of such faint stars are challenging to measure, because the faintest objects cannot be found in every individual exposure, but rather must be found by adding together exposures.  The stars were found by coadding the many GO-11677 exposures taken at nearly the same time in 2010, so the proper motions did not influence the finding.  To measure proper motions, one must examine images taken over a much larger period.
The target field has been observed more than 200 times between 2002 and 2008 with ACS and these are the images used to determine the positions.  While the 2010 data is very homogeneous, in terms of exposure time and field coverage, the early-epoch data is very inhomogeneous.  Typically there are one or two exposures at any given epoch, with exposure depths from 339s to 1500s (all archival images in filters other than F606W and F814W and with exposure times less than 339s were ignored).

Our analysis is a generalisation of previous studies of this type$^{35}$.
 We start by noting  that since the HST PSF is moderately
undersampled, much of the flux of a given star is concentrated in its
centermost pixel and if a star has any detectable impact on a given
exposure, it will push the flux in the centermost pixel above those of its eight surrounding
neighbors.  If it does not even do this, then we have no way of knowing
that it is there.  Of course, it takes many such marginal detections to be
certain that a star is there.  Our strategy of measuring the proper motion
will be to simply find the combination of position and motion
 that maximize the number of detections.

Since each peak in each exposure could correspond to a marginal detection,
we examine all of the available ACS exposures of the field
through F606W or F814W with at least 339s exposure (252 from program
GO-11677 in 2010 and 170 from other programs scattered between 2002 and
2012).  For each star, we identified the approximate position
 where the star would have been located in each exposure.  We then extracted a list of all
the local maxima within 3 pixels of this location and mapped the center of each
into the reference frame and tagged it with the time of the exposure.
Any of these peaks could correspond to a marginal detection of the object.
Typically there are 3 such potential detections in each exposure, and thus
about 1300 such detections over the full set of exposures.

We then took this list of 1300 events and did a simple grid search through
parameter space for the epoch-2010 position and proper motion that
returned the maximum number of detections.  We weighted each detection by
the exposure time of the image.  In this way, we determined the most
likely position and motion for each object.
This is not the most accurate way to measure motions for the relatively bright stars.  For those, it is far preferable to fit the PSF to each observation of the star, yielding a position good perhaps to 0.01 pixel, rather than the crude 0.7 pixel aperture used above. 
 However, when a faint local maximum is the only available information, our approach is optimal.

The Small Magellanic Cloud  moves at approximately (0.095,0.035) pixels per year, which amounts to a $\sim 0.5$~pixel shift over the roughly 5-year average baseline of the observations.  Even though a given peak can only yield a position to 0.7 pixel, by combining the information from many observations, one can clearly get proper motions to better than 0.1 pixel for the stars with many observations.  For the faintest stars, for which we have $\sim$ 50 observations at each epoch, we can still effectively separate the SMC from the cluster stars, even at the very bottom of the WDCS.

Figure~S2 shows the distribution of displacements as a function of F814W magnitude
for stellar-like objects in our field (we have excluded all objects that are
measurably extended). The lower panel shows stars that lie bluewards of a colour cut
defined by $F814W=24 + 3.25 \left( F606W-F814W \right)$, and the upper panel shows
the corresponding redder objects. We use the solid line
as our magnitude-dependant proper motion cut to isolate cluster members. The
resulting division into cluster and field populations is shown in Figure~S3. In
the left panel we have isolated a clear white dwarf sequence with essentially
no intrusion by the SMC main sequence. The right hand panel shows all non-cluster
members consistent with being stars. There is some leakage of the white
dwarf population into the field sample for $F606W>29$ and $F606W-F814W>1.2$, but
we account for this by modelling the distribution of proper motion displacements 
at each magnitude in order to incorporate this into our completeness. This is
shown in Figure~S3. Solid points show the distribution of proper motion
displacements for stars with $29<F606W<29.5$ and $F606W-F814W<1.2$. 
 The vertical dashed
line indicates the proper motion cut at the midpoint of this magnitude bin. In
all cases the correction to the number counts is $< 15\%$.

The fact that the true SMC main sequence extends well past the truncation of the
white dwarfs in Figure~S2 is a direct empirical demonstration that the white dwarf
luminosity function truncation is real.

\psfig{file=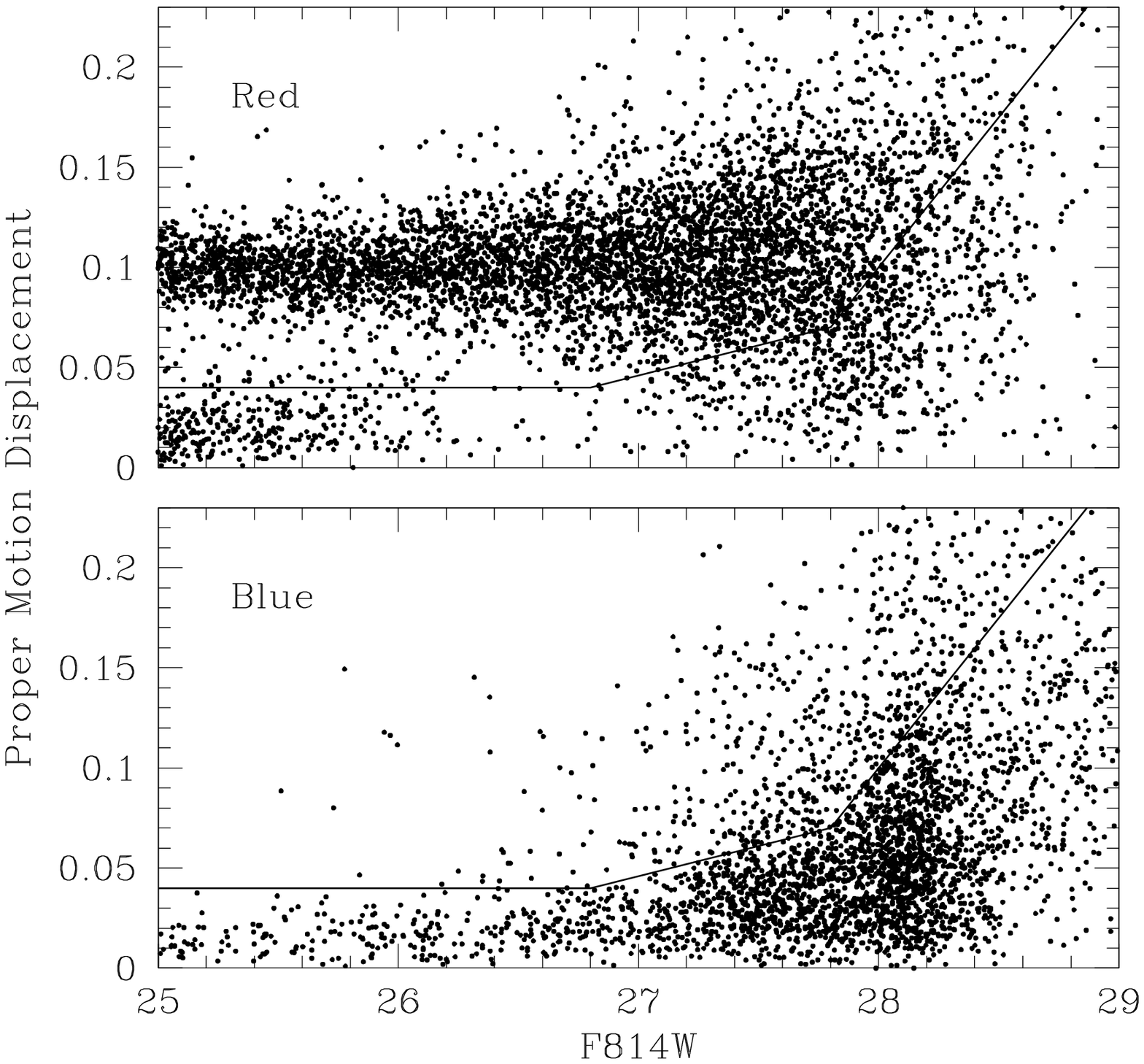,height=5in}

{\bf Figure~S2$\left| \right.$ Proper motion cut to separate cluster from background.} The positional displacement is the shift in
position between first and second epochs, relative to a standard of
rest defined by the bright cluster main sequence stars. Thus, a displacement
near zero indicates a cluster member. The upper panel shows the displacements
as a function of F814W magnitude for stars that lie redwards of a colour cut
$F814W=24 + 3.25 \left( F606W - F814W \right)$.
The lower panel shows the same quantity for stars that lie bluewards of this
cut. In both panels, the piecewise linear curve indicates the proper motion
cut we use to isolate cluster members.

\psfig{file=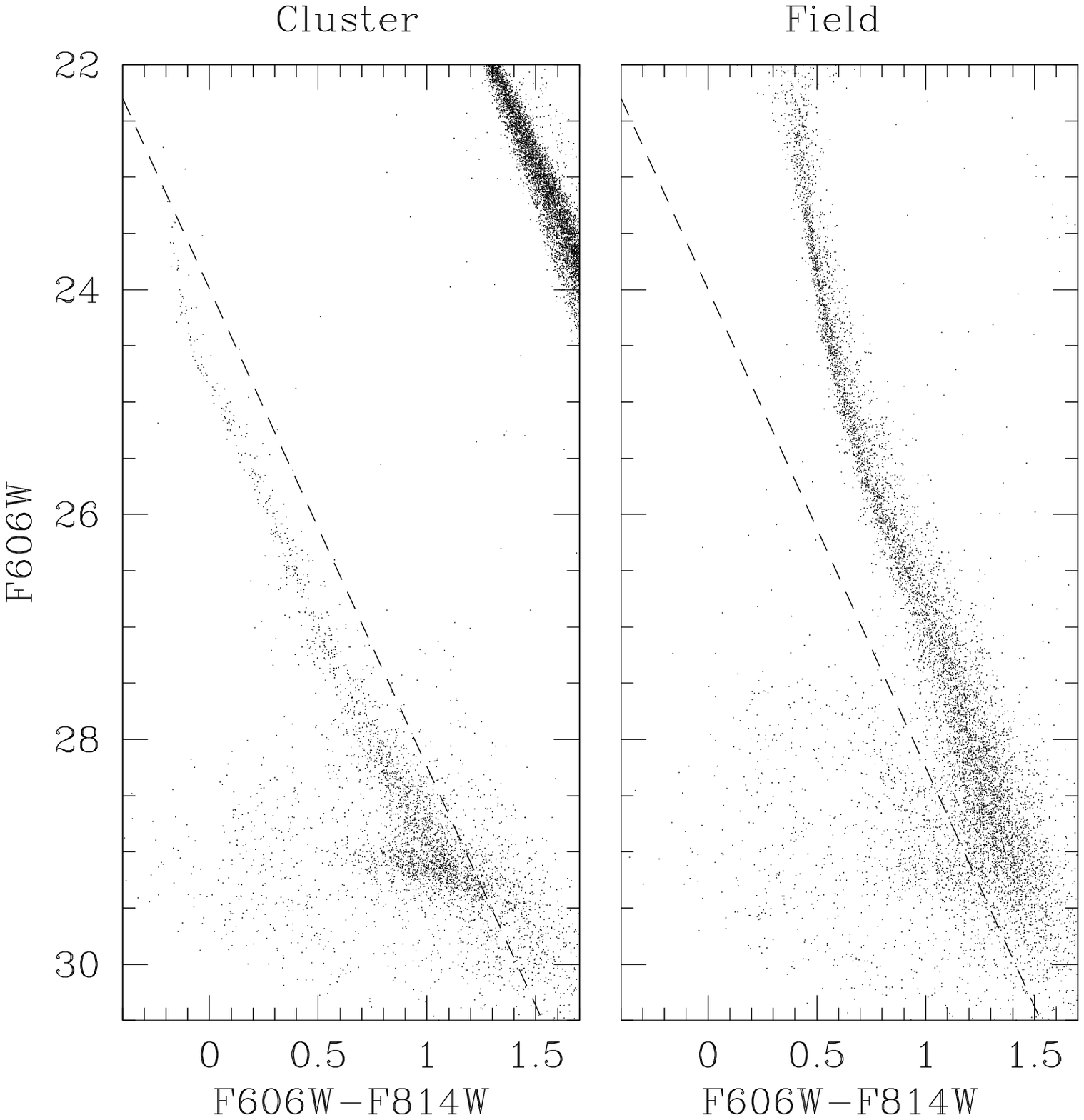,height=5in}

{\bf Figure~S3$\left| \right.$ Population Separation according to the proper motion cut.}
The left panel shows the cluster white dwarf cooling sequence isolated by the
proper motion cut in Figure~S2. The cluster main sequence can be seen towards the
upper right. The right hand panel shows the field population, dominated by the
main sequence of the Small Magellanic Cloud in the background. The field population
blueward of the white dwarf sequence is dominated by background galaxies unresolved
in our images and which therefore pass the stellarity cut. The slight bluewards extension of the
SMC main sequence at $F606W \sim 29$ is composed of cluster white dwarfs whose proper
motion errors bleed over our nominal cluster separation cut. We describe the correction
for this loss in Figure~S4. The dashed lines in both panels represents the colour cut
shown in Figure~S2.

\psfig{file=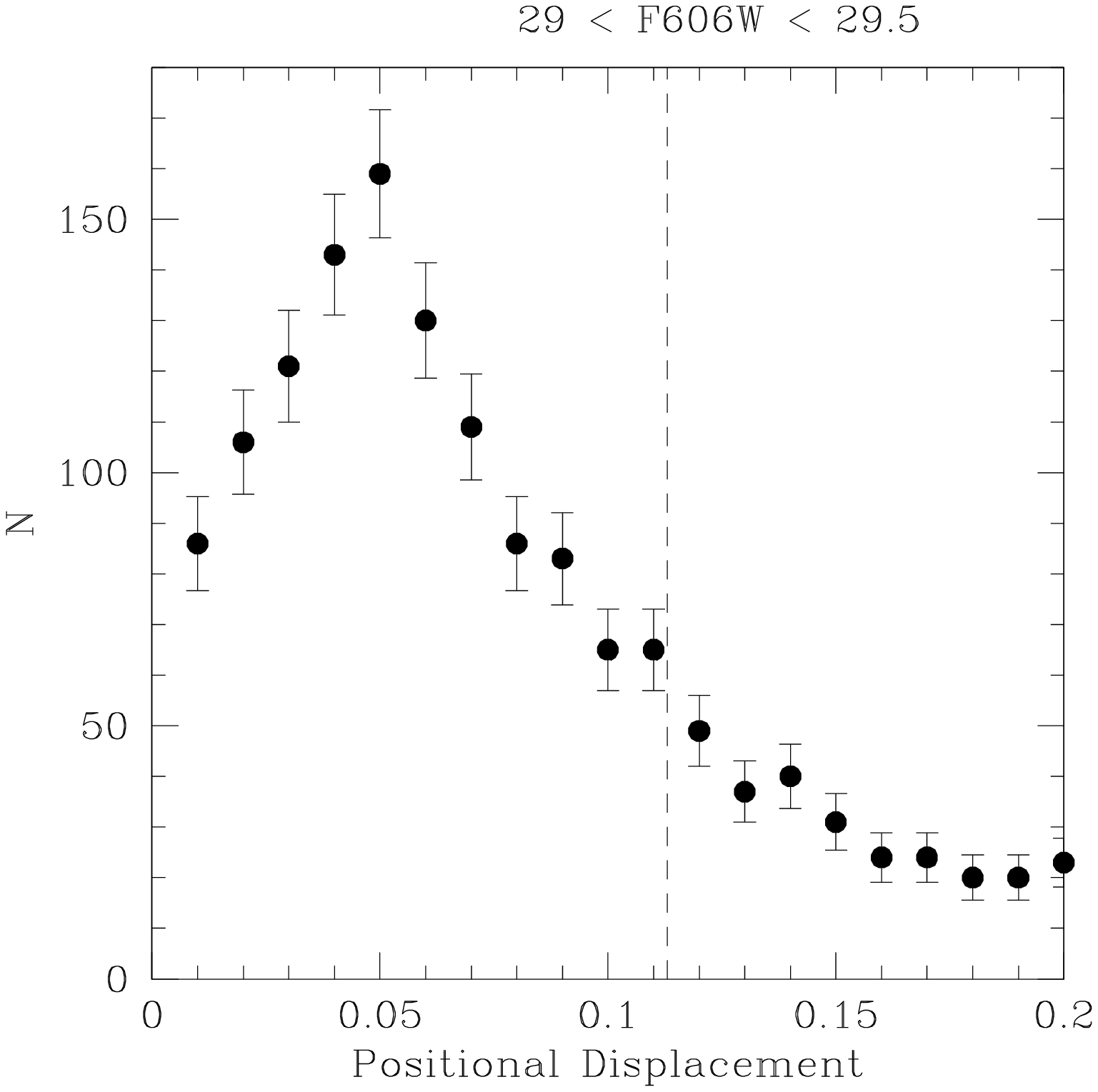,height=5in}

{\bf Figure~S4$\left| \right.$ Correcting for Lost White Dwarfs.}
The filled circles show the distribution of
proper motion displacements for all point sources
with $29<F606W<29.5$ and $F606W-F814W<1.2$ (i.e. blue
enough to be counted in the grid shown in Figure~1).
The vertical dashed line indicates the proper motion cut at
this magnitude. We account for the loss of white dwarfs across
this cut (as seen in Figure~3) by modelling the distribution
of proper motions at each magnitude, i.e. as shown by the solid points,
and including this in our completeness estimates.



\bigskip

{\bf The Distance to 47~Tucanae}
\bigskip

The determination of an absolute age ultimately rests on the determination of an apparent magnitude, no
matter which kind of star one is attempting to model, and therefore requires knowledge of the distance to
the cluster.
The fact that degenerate white dwarfs asymptote to a calculable radius that is independent of luminosity or effective
temperature makes them an attractive candidate for performing a distance determination. Furthermore, the fact that
the high gravity in white dwarfs drives gravitational sedimentation and produces pristine Hydrogen or Helium atmospheres
 means that white dwarf effective temperatures do not suffer from the systematic errors related to metallicity corrections
that one often faces with using lower gravity stars to measure a distance. However, the 
measurement of  a distance to 47~Tuc using the white dwarf cooling sequence$^{36}$,
 suggested a result markedly at odds with other determinations
and so a comprehensive re-examination of this issue is required.

Our goal is to provide a combination of model cooling sequence, distance and redenning that provides the best fit to
the data in our deep ACS field. This is different from the previous approach$^{36}$, where estimates are based
on comparing the target white dwarfs with a set of nearby calibrators observed in the same filters. The consequence is
that our distance is a function of the assumed white dwarf mass. 
We fit models of a given mass to the ACS data in the range $24.5 < F606W < 26$. We neglect brighter stars because the white dwarf
sequence shows a change of slope to the blue that suggests that the stars are still hot enough to retain a non-negligible contribution
to the radii from a non-degenerate equation of state. We neglect fainter stars because $F606W \sim 26$ corresponds to white dwarf
cooling ages $\sim 10^9$~years, and we wish to restrict our attention to a magnitude range over which the mass of the white dwarfs
is not likely to change significantly.
The effect of these cuts is that our fit spans the approximate effective temperature range $17000 K > T_{eff} > 8000K$.
 The resulting distance is $\mu_0=13.32 \pm 0.08$ for $E(B-V)=0.04\pm 0.02^{35}$ for a white dwarf mass of $0.53 M_{\odot}$ and
standard Hydrogen layer mass fraction $q_H=10^{-4}$.
 If we include, in quadrature, the error associated with the unknown model mass, assuming $0.53 \pm 0.03 M_{\odot}$,
we get a distance $\mu_0=13.32 \pm 0.09$.

This distance is in excellent agreement with the most recent determinations from main sequence eclipsing binaries$^{24,38}$
 and main sequence fitting$^{39}$, although somewhat shorter than older determinations which yielded $\mu_0 \sim 13.4$--13.5$^{6,40,41}$.
 It is larger than the $\mu_0=13.15 \pm 0.14$ previously determined from the white dwarf sequence, and
consistent with the value
$\mu_0=13.36 \pm 0.08$ determined by 
 fitting models to the data from the parallel observations taken for this
program$^{42}$.
The latter data were taken in the WF3 bandpasses, F390W, F606W and F110W, and the distance comes from a simultaneous fit to all
the bandpasses from blue to infra-red. Although better wavelength coverage offers stronger constraints on effective temperatures, a significant fraction
of the Woodley sample lie above our $T_{eff}$ limit and may thus be sensitive to corrections for non-degeneracy. Thus, we will adopt the number
based on the fit to the ACS data alone.

The source of the difference with the earlier result can be traced to an improved understanding of the calibrator population.
 In that work, a direct comparison was performed between the upper cooling sequence and
a series of local white dwarfs, with well-known parallaxes, in the HST instrumental bandpasses.
 This avoids the
need to use models to estimate the radii, but requires an empirical correction of the masses of the nearby
calibration sample to bring them into line with the anticipated mass of the cluster sequence.
The estimate of the correction due to this effect is $\delta \mu_0 = 2.4 ( <M_{wd}>-0.594 M_{\odot} )$,
using a mean mass for the calibrator sample determined from spectroscopic gravity measurements$^{36}$.
Recent measurements have claimed a somewhat higher mean mass for the solar neighbourhood white dwarfs when measured
from gravitational redshifts$^{43}$,
with a mean mass of $0.64 M_{\odot}$ for DA white dwarfs -- a difference of $\sim 0.07 M_{\odot}$ relative
to the same quantity determined by spectroscopic methods for a common sample$^{44}$. Similarly, recent improvements in modelling
of white dwarf spectra suggest an increase ($\sim 0.04 M_{\odot}$) in the mean mass estimate for the hot white dwarfs of the Palomar-Green sample$^{45}$. If an similar correction is required for the original calibrator sample, the above formula would suggest the revision of the original analysis would produce a
distance that lies in the range $\mu_0 \sim 13.26$--13.32, which would then be quite consistent with our estimate.

For the intercomparison of the two clusters NGC~6397 and 47~Tuc, the white dwarf cooling sequences can also be used
to perform a relative distance comparison, thereby avoiding some of the calibration issues associated with an absolute
distance determination. The left-hand panel of Figure~S5 shows the location of the two cooling sequences in 
apparent magnitude. In order to correct for the different redenning of the two clusters, we assume
that the change in slope on the upper cooling sequence (denoted by vertical dashed lines) occurs at the same absolute
colour. This in turn implies a corresponding extinction correction. Together these combine to imply a shift indicated
by the first (diagonal) green arrow. The difference in apparent magnitude is then an indication of the relative distance
between the two, represented by the second (vertical) green arrow. The resulting comparison (achieved by shifting the
NGC~6397 stars to the same scale as 47~Tuc) is shown in the right-hand panel. The differential numbers are 
$\Delta E(F606W-F814W)=0.19$, $\Delta A_{814}=0.35$ and $\Delta \mu_0 = 1.32 \pm 0.10$. We see that the
shifted NGC~6397 sequence extends to fainter magnitudes than the 47~Tuc sequence, --  another demonstration of the 
difference in cluster ages. This differential distance is also consistent with our absolute distance estimates
for the two clusters, $\Delta \mu_0 = 13.32 - 12.03 = 1.29 \pm 0.09$.

\psfig{file=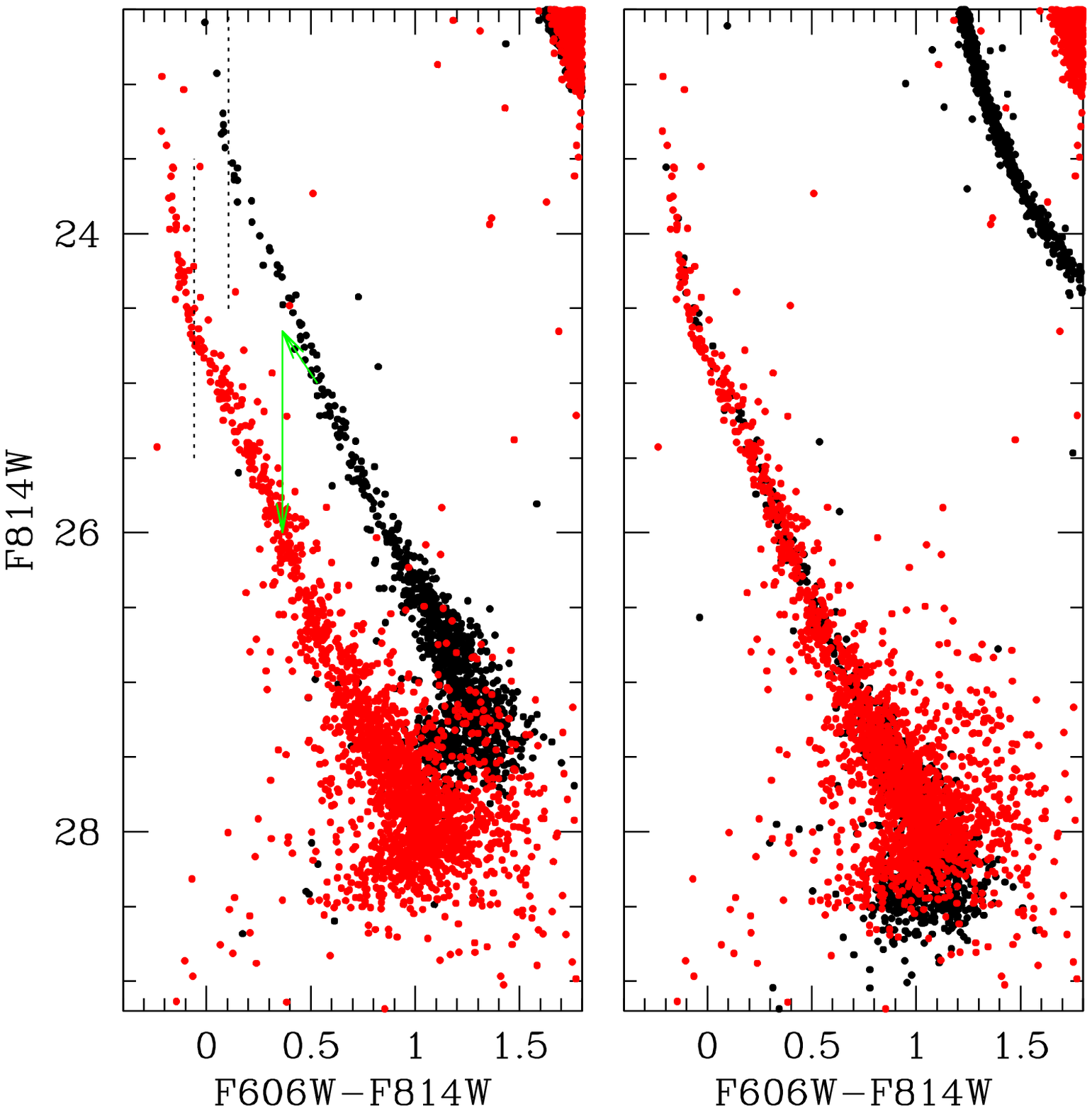,height=5in}
{\bf Figure~S5$\left| \right.$ Differential Comparison of the White Dwarf Sequences in 47~Tuc and NGC~6397.}
In each panel, the black points are proper-motion selected members of the NGC~6397 cluster,
and the red points are proper-motion selected members of 47~Tucanae. In the left hand panel,
the observed values for both clusters are shown. The vertical dotted lines indicate the change
in sequence slope that we use to determine differential redenning and the green arrows indicate
the resulting shifts used to correct for this and the different distances, in order to bring the
NGC~6397 stars onto the same distance scale as 47~Tuc. The resulting comparison is shown in the
right hand panel. The two populations closely follow the same sequence, although the NGC~6397 white
dwarfs extend to fainter magnitudes, as befits the larger cluster age. We note also that the main
sequences (in the upper right hand corner) do not lie on top of one another, because of the different
cluster metallicities.

\medskip

{\bf The effect of Metallicity on White Dwarf Ages}

\bigskip

One of the attractions of using white dwarfs to determine the age of a stellar population is that their
high gravities cause heavier elements to settle out of the atmosphere. Thus, the conversion of colours to
effective temperatures should be much less sensitive to the metallicity of the population than in the
case of main sequence turnoff stars. 

On the other hand, stars of different metallicities will experience different core temperatures and therefore
potentially a difference in the ratio of Carbon to Oxygen in the core. In order to calibrate the sensitivity
of white dwarf cooling, we have calculated new models using progenitors of appropriate
metallicity for 47~Tuc. Using the MESA code$^{17}$, we have calculated the evolution of solar-mass stars
with $\left[ Fe/H \right] = -0.75$ and $\left[ \alpha / Fe \right]=+0.2$. We have also verified that these produce similar
results to the models of Dotter et al.$^{19}$, in order to facilitate comparison with analysis of the main sequence.
In addition, many globular clusters are now known to exhibit a spread in Helium abundance$^{46}$, and so we also
examined the observed$^{18}$ range of Y=0.25--0.27 for 47~Tuc.
 The
principal effect of these different progenitor histories is to change the relative fraction and distribution of Carbon and
Oxygen in the cores of white dwarfs, which can in turn affect the heat capacity of the white dwarf core, as well as determine
the amount of energy released if gravitational separation occurs upon crystallisation. The C/O profile can also be affected
by assumptions regarding the kind of mixing that occurs during the progenitor evolution.

 Figure~S6 shows the Oxygen
mass fraction profiles for four white dwarfs with masses in the range 0.50--0.53$M_{\odot}$. The black curve is the profile
for our standard models, based on evolutionary calculations$^{47}$. These are the models used for the age determination
in NGC~6397$^{12}$.
The blue curve represents the result of a Y=0.25, 
47~Tuc metallicity model in which convection is calculated assuming the standard Schwarzschild convection. The red curve
shows the result for the same progenitor model, but the model is also allowed to convectively mix according to the Ledoux
criterion (i.e. convection driven by chemical-composition gradients). Convection affects mostly the central Oxygen concentration.
Finally, the green model is the same as the blue curve, except that Y=0.27.

\psfig{file=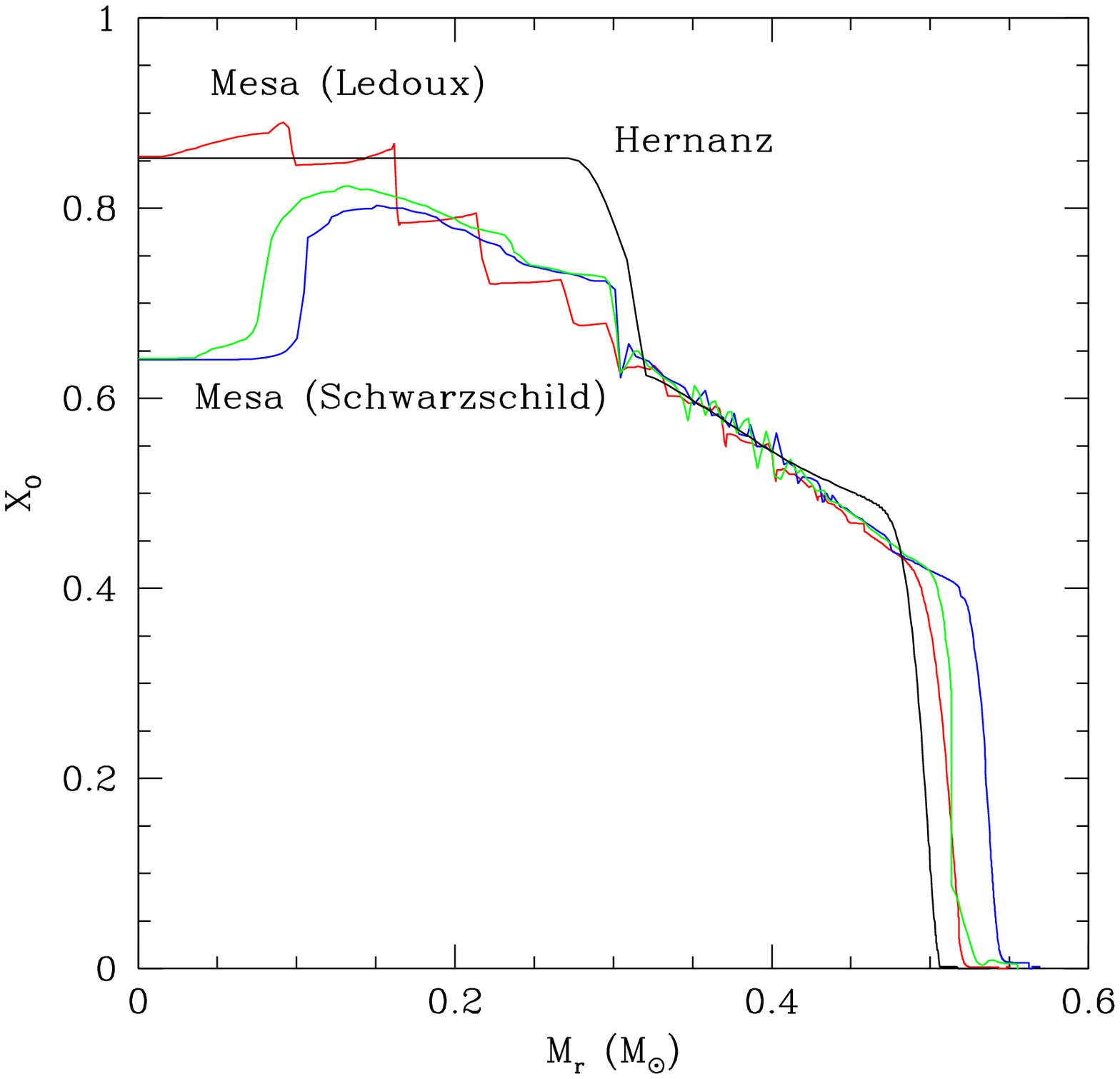,height=5in}

{\bf Figure~S6$\left| \right.$ Metallicity Effect on White Dwarf Core Composition.}
The four curves show the radial profile of the Oxygen mass fraction
in a $0.53 M_{\odot}$ white dwarf under three different progenitor
models. The black line is based on the models$^{47}$ we use
as the basis for the age determination in NGC6397$^{12}$. The
red and blue profiles are compositional profiles calculated for
progenitors of the metallicity of 47~Tuc. The differences result
from different prescriptions for convective mixing, depending on
whether (red) or not (blue) one allows for convection driven by
compositional gradients (Ledoux convection). The green curve is
the same as the blue curve, but the initial Helium mass fraction
was Y=0.27, rather than Y=0.25.
The resulting differences in these models 
do not lead to significant changes in the cooling age.

Fitting the Schwarzschild models to the Hess distribution results in an age of $9.9 \pm 0.7$~Gyr (95\% confidence), while
the models with Ledoux convection result in an age of $9.8 \pm 0.4$~Gyr. This is similar to our standard models, which
is perhaps not surprising as they have similar central Oxygen as do our standard models (Figure~S5). The progenitor
models assumed Y=0.25 in the aforementioned cases, and we have also calculated equivalent models with Y=0.27. The resulting
core profiles are very similar and result in no significant change in the age determination.

{\bf Supplementary References}

{\setlength{\parindent}{0cm}

31. Fruchter, A. S. \& Hook, R. N., A Method for the Linear Reconstruction of Undersampled Images., 2002, {\it PASP}, {\bf 114}, 144

32. Stetson, P. B., DAOPHOT -- A computer program for crowded-field stellar photometry., 1987, {\it PASP}, {\bf 99}, 191

33. Stetson, P. B., The center of the core-cusp globular cluster M15: CFHT and HST Observations, ALLFRAME reductions., 1994, {\it PASP}, {\bf 106}, 250 

34. Kalirai, J. S. et al.,A Deep, Wide-field, and Panchromatic View of 47 Tuc and the SMC with HST: Observations and Data Analysis Methods., 2012, {\it AJ}, {\bf 143}, 11

35. Anderson, J., et al., 
Deep Advanced Camera for Surveys Imaging in the Globular Cluster NGC 6397: Reduction Methods, 2008, {\it AJ}, {\bf 135}, 2114 

36. Zoccali, M., et al., The White Dwarf Distance to the Globular Cluster 47 Tucanae and its Age, 2001, {\it ApJ}, {\bf 553}, 733 

37. Harris, W. E., A Catalog of Parameters for Globular Clusters in the Milky Way, 1996, {\it AJ}, {\bf 112}, 1487 

38. Kaluzny, J., et al.,
The Clusters Ages Experiment (CASE). II. The Eclipsing Blue Straggler OGLEGC 228 in the Globular Cluster 47 Tuc, 2007, {\it AJ}, {\bf 134}, 541 

39. Bergbusch, P. A. \& Stetson, P. B., A New Color-Magnitude Diagram for 47 Tucanae: A Statistical Analysis, 2009, {\it AJ}, {\bf 138}, 1455 

40. Carretta, E., Gratton, R. G., Clementini, G., \& Fusi Pecci, F., Distances, Ages, and Epoch of Formation of Globular Clusters, 2000, {\it ApJ}, {\bf 533}, 215 

41. Reid, N., HIPPARCOS subdwarf parallaxes - Metal-rich clusters and the thick disk, 1998, {\it AJ}, {\bf 115}, 204

42. Woodley, K. A., et al., The Spectral Energy Distributions of White Dwarfs in 47 Tucanae: The Distance to the Cluster, 2012, {\it AJ}, {\bf 143}, 50 

43. Falcon, R. E., Winget, D. E., Montgometry, M. H., \& Williams, K. A., A Gravitational Redshift Determination of the Mean Mass of White Dwarfs. DA stars, 2010, {\it ApJ}, {\bf 712}, 585 
	
44. Koester, D., et al., High-resolution UVES/VLT spectra of white dwarfs observed for the ESO SN Ia Progenitor Survey. III. DA white dwarfs, 2009, {\it A\&A}, {\bf 505}, 441 

45. Tremblay, P.-E. \& Bergeron, P., Spectroscopic Analysis of DA White Dwarfs: Stark Broadening of Hydrogen Lines Including Nonideal Effects, 2009, {\it ApJ}, {\bf 696}, 1755 

46. Piotto, G., Observations of multiple populations in star clusters, 2009, in IAU Symposium 258, {\it The Ages of Stars}, ed. E. E. Mamajek, D. R. Soderblom \& R. G. Wyse, (Cambridge: Cambridge University Press), 233

47.  Hernanz, M., et al., The influence of crystallization on the luminosity function of white dwarfs, 1994, {\it ApJ}, {\bf 434}, 652
}